\documentstyle[12pt,a4,fleqn,epsfig]{article}
\normalbaselineskip=\baselineskip

\def    \bc            {\begin{center}}
\def    \ec            {\end{center}}
\def    \DD            {{\cal D}}
\def    \LL            {{\cal L}}
\def    \dmu           {\partial_\mu}
\def    \dnu           {\partial_\nu}
\def    \dd            {\displaystyle}
\def    \be            {\begin{equation}}
\def    \ee            {\end{equation}}
\def    \bea           {\begin{eqnarray}}
\def    \eea           {\end{eqnarray}}
\def    \nn            {\nonumber}
\def    \Asm           {\mbox{$ A_{\lambda{\bar\lambda}}^{SM} $}}

\def    \DAg           {\mbox{$ \Delta A_{\lambda{\bar\lambda}}^\gamma $}}
\def    \DAz           {\mbox{$ \Delta A_{\lambda{\bar\lambda}}^Z $}}
\def    \M2            {\mbox{$ \tilde{\cal M} $}}
\def    \Mg            {\mbox{$ {\tilde{\cal M}}^\gamma $}}
\def    \Mz            {\mbox{$ {\tilde{\cal M}}^Z $}}
\def    \Mnu           {\mbox{$ {\tilde{\cal M}}^\nu $}}

\def    \raw           {\rightarrow}

\def    \lraw          {\leftrightarrow}

\begin{document}
%
%
\begin{titlepage}
\vskip 3cm
\begin{center}
{\Large\bf Heavy Fermions Virtual Effects at $e^+ e^-$ Colliders\footnote{
Work supported in part by the European Union
under contract No.~CHRX-CT92-0004, and by MURST.}
}
\end{center}
\vskip 3.0cm

\begin{center}
{\large Stefano Rigolin }
\\
\vskip .1cm
Dipartimento di Fisica, Universit\'a di Padova, Italy
\\
INFN, Sezione di Padova, Padova, Italy
\\
E-Mail Address: RIGOLIN@VAXFPD.PD.INFN.IT
\\
\vskip .2cm
\end{center}

\vskip 4cm

\begin{abstract}
\noindent
We derive the low-energy electroweak effective lagrangian for the case 
of additional heavy, unmixed, sequential fermions.
Present LEP1 data still allow for the presence of a new quark and/or lepton
doublet with masses greater than $M_Z/2$, provided that these multiplets 
are sufficiently degenerate.
Keeping in mind this constraint, we analyse the virtual effects of 
heavy chiral fermions in $e^+ e^- \raw W^+ W^-$ process at energies 
$\sqrt{p^2} = 200,~500,~1000$ GeV, provided by LEP2 and Next Linear Collider.
The effects will be unobservable at LEP2, being smaller than $3.0\cdot 10^{-3}$,
while more interesting is the case of NLC, where an enhancement factor, 
due to a delay of unitarity, gives deviations from SM of the order 10-50 per 
cent, for a wide range of new fermions masses.
\end{abstract}

\vfill{
\noindent
Talk given at the XIX International School of Theoretical Physics: 
"Particle Physics and Astrophysics in the Standard Models and Beyond",
Bystra, 19-26 September 1995.}

\end{titlepage}
\setcounter{footnote}{0}


\section{Introduction}

LEP1 precision data represent a step of paramount relevance in probing 
extensions of the Standard Model (SM).
Through their virtual effects, the electroweak radiative corrections "feel"
the presence of new particles running in the loops and the level of
accuracy on the relevant observables is such that this set of tests is
complementary to the traditional probes on virtual effects due to new
physics (i.e. highly suppressed or forbidden flavour changing neutral
current phenomena). In some cases, as that which we aim to discuss here,
the electroweak precision tests represent the only indirect way to
search for these new particles.

We will discuss electroweak radiative effects from extensions of the 
ordinary fermionic spectrum of the SM.
The new fermions are supposed to possess the same colour and electroweak 
quantum numbers as the ordinary ones and to mix very tinily with the 
ordinary three generations.
The most straightforward realization of such a fermionic extension
of the SM is the introduction of a fourth generation of fermions. 
This possibility has been almost entirely jeopardized
by the LEP1 bound on the numbers of neutrinos species. Although
there still exists the obvious way out of having new fermion
generations with heavy neutrinos, we think
that these options are awkward enough not to deserve further studies.
Rather, what we have in mind in tackling this problem are
general frames discussing new physics beyond the SM
which lead to new quarks and/or leptons classified in
the usual chiral way with iso-doublets and iso-singolets for
different chiralities.
Situations of this kind may be encountered in grand
unified schemes where the ordinary fifteen Weyl spinors of each
fermionic generation are only part of larger representations or
where new fermions (possibly also mirror fermions) are requested
by the group or manifold structure of the schemes.
Chiral fermions with heavy static masses may also provide a first 
approximation of virtual effects in techicolor-like schemes when 
the dynamical behaviour of technifermion self-energies are neglected.

Although such effects have been extensively investigated in the literature
\cite{gen}, our presentation focuses mainly on two aspects, which
have been only partially touched in the previous analyses: the use
of effective lagrangians for a model-independent treatment of the problem
and a discussion of the validity of this approach in comparison with the
computation in the full-fledged theory.

While separate tests can be set up for each different extension of the SM,
there may be some advantage in realizing this analysis in a model independent
framework. The natural theoretical tool to this purpose
is represented by an effective electroweak lagrangian where,
giving up the renormalizability requirement, all $SU(2)_L\otimes U(1)_Y$ 
invariant operators up to a given dimension are present with unknown 
coefficients, 
to be eventually determined from the experiments. Each different model 
fixes uniquely this set of coefficients and the effective lagrangian becomes 
in this way a common ground to discuss and compare several SM extensions.
The introduction of the well known $S$, $T$ and $U$ \cite{stu} or
$\epsilon$'s \cite{eps} variables was much in the same spirit and the 
use of an effective lagrangian represents in a sense the natural extension 
of these approaches (section 2).

We derive (section 3) some constrains on new chiral doublets, from latest 
avaible values of the $\epsilon$'s data, in the effective lagrangians 
approximation 
and in the full one-loop computation, putting on evidence that deviations are 
sizeable (compared with experimental errors) only for fermion masses close to 
the $M_Z/2$ threshold.
Some of the constraints on new sequential fermions coming from $p \bar{p}$
accelerator results are also presented.

But LEP1 analysis is not important only for studying the existence of new 
physics in the bilinear sector. Also trilinear coupling are severely 
constrained by the presence of observed bilinear effects at LEP1-SLC.
To avoid ambiguities in the forms factor definitions at one loop level, we 
present an analysis of new chiral fermions effects on $e^+ e^- \raw 
W^+ W^-$ differential cross-section. Here it will be evident that the 
delay's of unitarity effects makes higher energies collider ($\sqrt{p^2}=500$ 
GeV, or $1000$ GeV, much more sensitive than LEP2 to this kind of new physics 
effects (section 4).

\section{Effective Lagrangian Approach}

The use of an effective lagrangian for the electroweak physics
has been originally advocated for the study of the large Higgs mass limit
in the SM \cite{abe,alo,her}. Subsequently, contributions from chiral 
$SU(2)_L$ doublets have been considered in the degenerate case
\cite{dho}, for small splitting \cite{app} and in the case of infinite splitting
\cite{ste,fmm}.
In the present note we will deal with the general case
of arbitrary splitting among the fermions in the doublet 
\cite{fmrs}.

Here, for completeness, we consider the standard list of 
$SU(2)_L\otimes U(1)_Y$ 
CP conserving operators containing up to four derivatives and built out of the 
gauge vector bosons $W^i_\mu~~(i=1,2,3),~B_\mu$ and the would be Goldstone 
bosons $\xi^i$ \cite{alo}:
\bea
\LL_0&=&\dd\frac{v^2}{4}[tr(TV_\mu)]^2\nn\\
\LL_1&=&i\dd\frac{gg'}{2}B_{\mu\nu}tr(T\hat W^{\mu\nu})\nn\\
\LL_2&=&i\dd\frac{g'}{2}B_{\mu\nu}tr(T[V^\mu,V^\nu])\nn\\
\LL_3&=&g~tr(\hat W_{\mu\nu}[V^\mu,V^\nu])\nn\\
\LL_4&=&[tr(V_\mu V_\nu)]^2\nn\\
\LL_5&=&[tr(V_\mu V^\mu)]^2\nn\\
\LL_6&=&tr(V_\mu V_\nu)tr(TV^\mu)tr(TV^\nu)\nn\\
\LL_7&=&tr(V_\mu V^\mu)[tr(TV^\nu)]^2\nn\\
\LL_8&=&\dd\frac{g^2}{4}[tr(T\hat W_{\mu\nu})]^2\nn\\
\LL_9&=&\dd\frac{g}{2}tr(T\hat W_{\mu\nu})tr(T[V^\mu,V^\nu])\nn\\
\LL_{10}&=&[tr(TV_\mu)tr(TV_\nu)]^2\nn\\
\LL_{11}&=&tr((\DD_\mu V^\mu)^2)\nn\\
\LL_{12}&=&tr(T\DD_\mu \DD_\nu V^\nu)tr(TV^\mu)\nn\\
\LL_{13}&=&\dd\frac{1}{2}[tr(T \DD_\mu V_\nu)]^2\nn\\
\LL_{14}&=&i~g~\epsilon^{\mu\nu\rho\sigma}~tr(\hat W_{\mu\nu}V_\rho)
tr(TV_\sigma)
\label{a2}
\eea
We recall the notation used. If we define the Goldstone boson contribution 
$ U = exp(i \vec\xi\cdot\vec\tau / v)$ (so that in the unitary gauge $U = 1$),
than:
\be
T = U\tau^3U^\dagger~~~,~~~V_\mu = (D_\mu U) U^\dagger~~~,
\label{a3}
\ee
\be
D_\mu U = \dmu U - g \hat W_\mu U+g'U\hat B_\mu~~~,
\label{a5}
\ee
\noindent
where $\hat W_\mu,~\hat B_\mu$ are the matrices collecting the gauge fields:
\be
\hat W_\mu = \frac{1}{2i}\vec W_\mu\cdot\vec\tau~~~,~~~
\hat B_\mu = \frac{1}{2i} B_\mu\tau^3~~~.
\label{a6}
\ee

\noindent
The corresponding field strengths are given by:
\bea
\hat W_{\mu\nu}&=&\dmu\hat W_\nu-\dnu\hat W_\mu-g[\hat W_\mu,\hat W_\nu]
{}~~~,\nn\\
\hat B_{\mu\nu}&=&\dmu\hat B_\nu-\dnu\hat B_\mu~~~.
\label{a7}
\eea
\noindent
Finally the covariant derivative acting on $V_\mu$ is given by:
\be
\DD_\mu V_\nu=\dmu V_\nu-g[\hat W_\mu,V_\nu]~~~.
\label{a8}
\ee
\noindent
The effective electroweak lagrangian reads:
\be
\LL_{eff}=\LL_{SM} + \sum_{i=0}^{14} a_i \LL_i~~~,
\label{a9}
\ee
\noindent
where $\LL_{SM}$ is the "low-energy" SM lagrangian, and all the contributions 
of the new physics heavy sectors is contained in the $a_i$ coefficients 
\footnote{Here we do not need to include the Wess-Zumino term \cite{fmm}.}.

We have determined the coefficients $a_i~(i=0,...,14)$, for an extra doublet 
of heavy fermions (quarks or leptons), by computing the corresponding one-loop 
contribution to a set of $n$-point gauge boson functions $(n=2,3,4)$, in the 
limit of low external momenta.
For example just look at the two-point vector boson functions 
$-i\Pi^{\mu\nu}_{ij}(p)$.
In the limit $ p^2 \ll 4 M^2 $, we can use a derivative expansion around 
$p^2=0$: 
\bea
\Pi^{\mu\nu}_{ij} (p) & = & g^{\mu\nu} \Pi_{ij}(p^2)+(p^\mu p^\nu ~~{\rm terms})
~~~~~~~~~~~~~~~~~~~~~~~(i,j=0,1,2,3) \nn \\
\Pi_{ij}(p^2) & \equiv & A_{ij}+p^2 F_{ij}(p^2)= A_{ij} + p^2 F_{ij}(0)+ ...
\label{a0}
\eea
The next terms in the $p^2$ expansion are suppressed by increasing
powers of $p^2/M^2$, ($M$ generically representing the mass of the particles
running in the loop), and so will be neglected in this effective lagrangian 
approach. By denoting with $M$ and $m$ the masses of the upper and lower
weak isospin components and with $r=m^2/M^2$ the square ratio, we obtain, in 
units of $1/16\pi^2$:
\bea
a^q_0  &=&  {3 M^2 \over 2 v^2}\left( { 1 - r^2 + 2\,r\,\log r
\over 1-r} \right)
\nn\\
a^q_1  &=& \frac{1}{12 (-1+r)^3}\left[ {3(1 - 15 r + 15 r^2 - r^3) + 2 (1 -
12 r  -6 r^2  - r^3 )\log r } \right]
\nn\\
a^q_2 &=& \frac{1}{12 (-1+r)^3}\left[{3(3 - 7\,r + 5\,{r^2} - r^3)+ 2\,(1 -
{r^3})\,\log r } \right]
\nn\\
a^q_3 &=& \frac{1}{8 (-1+r)^3}\left[ 3 ( -1 + 7\,r - 7\,{r^2} +
{r^3}) + 6\,r\,(1 + {r})\,\log r) \right]
\nn\\
a^q_4 &=& \frac{1}{6 (-1+r)^3}\left[ 5 - 9\,r + 9\,{r^2} - 5\,{r^3} +
3\,(1 + {r^3})\,\log r \right]
\nn\\
a^q_5 &=& \frac{1}{24 (-1+r)^3}\left[ -23 + 45\,r - 45\,{r^2} + 23\,{r^3}
- 12\,(1+{r^3})\,\log r \right]
\nn\\
a^q_6  &=& \frac{1}{24 (-1+r)^3}
\left[ -23 + 81\,r - 81\,r^2 + 23\,r^3
-6\,(2 -3\,r\, -3\,r^2\,
+2\,r^3)\,\log r \right]
\nn\\
a^q_7  &=&-a^q_6
\nn\\
a^q_8 &=&\frac{1}{12 (-1+r)^3}\left[ 7 - 81\,r + 81\,r^2 - 7\,r^3
+ 6\,(1 - 6\,r\, - 6\,r^2\, + r^3)\,\log r \right]
\nn\\
a^q_9 &=&-a^q_6
\nn\\
a^q_{10}&=&0
\nn\\
a^q_{11}&=&-{1\over{2}}
\nn\\
a^q_{12}&=&\frac{1}{8 (-1+r)^3}
\left[1 + 9\,r - 9\,{r^2} - {r^3} + 6\,r\, (1+
{r})\,\log r\right]
\nn\\
a^q_{13} &=& 2\,a^q_{12}
\nn\\
a^q_{14} &=&\frac{3}{8 (-1+r)^2}\left[ 1 - r^2 + 2 r \log r
\right]
\label{b1}
\eea
\noindent
for quarks, and:
\bea
a^l_i  &=&\frac{1}{3} a^q_i~~~~~~~~~(i=0,~~i=3,...14)
\nn\\
a^l_1  &=&  \frac{1}{ 12 (-1+ r)^3}
\left[ 1 - 15  r + 15  r^2 -  r^3 - 2\,(1 + 6  r^2 - r^3 )\log  r  \right]
\nn\\
a^l_2 &=&\frac{1}{ 12 (-1+ r)^3}
\left[-1 - 3\, r + 9 r^2 - 5 r^3- 2\,(1 - r^3)\,\log  r  \right]
\label{b2}
\eea
\noindent
for leptons.

Indeed the use of an effective lagrangian in precision tests has its own
limitations. One can ask how large has to be $M$ to obtain a sensible 
approximation from the truncation of the full one-loop result. 
We will see this aspects in the next section.


\section{Two Point Functions}

For new chiral fermions which do not mix with the ordinary ones, the virtual 
effects measurable at LEP1 are all described by operators bilinear in the 
gauge vector bosons. We will describe these effects in the approximate 
effective theory (we can call it with evident meaning "static approximation"), 
as well in the full one-loop calculation. 

\subsection{Static Approximation}

The coefficients $a_i$ of the effective lagrangian $\LL_{eff}$ are related
to measurable parameters. In particular, to make contact with the LEP1 data, 
we recall that, by neglecting higher derivatives, the relation between the 
effective lagrangian $\LL_{eff}$ and the $\epsilon_i$ parameters, is given by:
\bea
\Delta\epsilon_1&=&2 a_0~,\nn\\
\Delta\epsilon_2&=&-g^2(a_8+a_{13})~,\nn\\
\Delta\epsilon_3&=&-g^2(a_1+a_{13})~.
\label{a11}
\eea
where $\Delta\epsilon_i$ are the new physics contributions to the $\epsilon$'s.
From eqs. (\ref{b1}) and (\ref{b2}) one finds:
\be
\Delta\epsilon^q_1=3 \Delta\epsilon^l_1=
{3 M^2 \over 8 \pi^2 }{G \over \sqrt{2}} \left[ { 1 - r^2
+ 2\,r\,\log r \over (1-r)} \right]
\label{a12}
\ee
\be
\Delta\epsilon^l_2=3 \Delta\epsilon^l_2 =
 {G m_W^2 \over 12 \pi^2 \sqrt{2}} \left[ {5 - 27r + 27
r^2-5 r^3+\left(3 - 9 r - 9 r^2 +3 r^3 \right) \log r \over \left( 1 - r
\right)^3} \right]
\label{a13}
\ee
\be
\Delta\epsilon^q_3 = {G m_W^2 \over 12 \pi^2 \sqrt{2}} \left[ 3 +  \log r
\right]
\label{a14}
\ee
\be
\Delta\epsilon^l_3 = {G m_W^2 \over 12 \pi^2 \sqrt{2}} \left[ 1 - \log r
\right]
\label{a15}
\ee

The $\epsilon_i$ parameters are obtained by adding to $\Delta\epsilon_i$
the SM contribution $\epsilon^{SM}_i$, which we regard as functions of
the Higgs and top quark masses.
A recent analysis of the available precision data from LEP1, SLD,
low-energy neutrino scatterings and atomic parity violation experiments,
leads to the following values for the $\epsilon_i$ parameters \cite{fit}:
\bea                                                         
\epsilon_1&=&(3.6\pm 1.5)\cdot 10^{-3}\nn\\
\epsilon_2&=&(-5.8\pm 4.3)\cdot 10^{-3}\nn\\
\epsilon_3&=&(3.6\pm 1.5)\cdot 10^{-3}
\label{eps}
\eea
Notice the relatively large error in the determination of $\epsilon_2$,
mainly dominated by the uncertainty on the $W$ mass.

It is clear from eq. (\ref{b1}) and from eqs. (\ref{a11}--\ref{a15}) 
that only $\Delta\epsilon_1$ can have a huge contribution proportional to $M^2$.
But, as it is well known, this term is vanishing in the limit of degenerate
doublet. 
From the analysis of this parameter we can obtain only a limitation of the 
splitting of the fermion masses, and not an "absolute" statement on the number
of possible extra doublets.  
In fig.1 we can see that if for relatively light masses $M=200$ GeV 
(dotted line) a small splitting is still allowed, ($ 0.5 \le r \le 1.5$), 
for heavier masses, like for example $M=1000$ GeV (full line), the doublet 
must be practically degenerate ($ 0.91 \le r \le 1.08$).

\begin{figure}[t]
\vspace{0.1cm}
\centerline{
\epsfig{figure=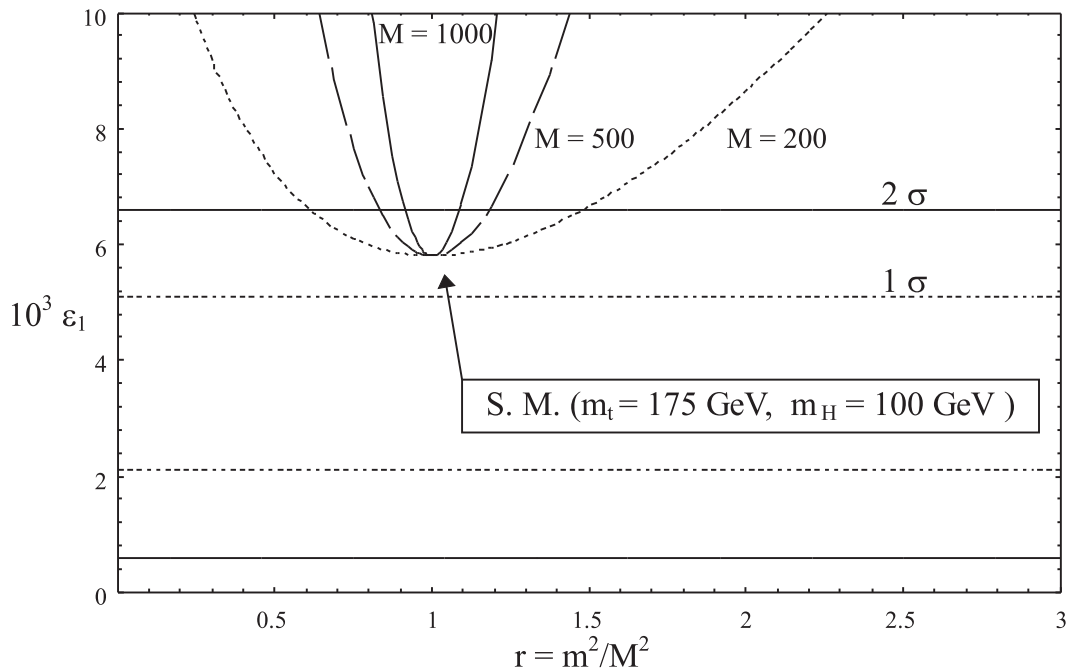,height=9cm,angle=0}}
\caption{\footnotesize\it 
Predictions for $\epsilon_1$ from an additional heavy quark doublet with masses 
$M=200$ {\rm GeV} (dotted line), $M=500$ {\rm GeV} (dashed line), $M=1000$
{\rm GeV} (full line), for $m_t=175$ {\rm GeV} and $M_H=100$ {\rm GeV}.
The $1 \sigma$ (dotted horizontal line) and $2 \sigma$ (full horizontal line) 
allowed region are also displayed. The values $\epsilon_1(r=1)$ is the SM 
predictions for the fixed $m_t=175$ {\rm GeV} and $M_H=100$ {\rm GeV}.}
\label{pol0.eps}
\end{figure}

The contributions from $\Delta\epsilon_2$ and $\Delta\epsilon_3$, have only 
dependence from $\log r$ and powers of $r$. Again $\Delta\epsilon_2$ is 
vanishing for $r=1$, so doesn't give us any useful indications. 
Different is the case of $\Delta\epsilon_3$, the only bilinear parameter non
null in the $r \raw 1$ limit. From eq. (\ref{a15}) we have:
\be
\Delta\epsilon^q_3 = 3 \Delta\epsilon^l_3 = {G m_W^2 \over 4 \pi^2 \sqrt{2}}
\simeq 1.3 \cdot 10^{-3} 
\ee
Thus, with the conditions provided us by the analysis of both $\epsilon_1$ and 
$\epsilon_3$, we obtain an "absolute" bound on the number of possible extra 
heavy fermions. 
The full horizontal lines in fig.2 are the $2 \sigma$ deviation from the
experimental value of $\epsilon_3$. It can be noted that at least one quark
doublet\footnote{Also one complete extra generation is allowed. 
In fact the leptonic contribution in the $r=1$ limit is just $1/3$ the 
hadronic contribution.} (full line) is not completely ruled out by the 
experiment.

\begin{figure}[t]
\vspace{0.1cm}
\centerline{
\epsfig{figure=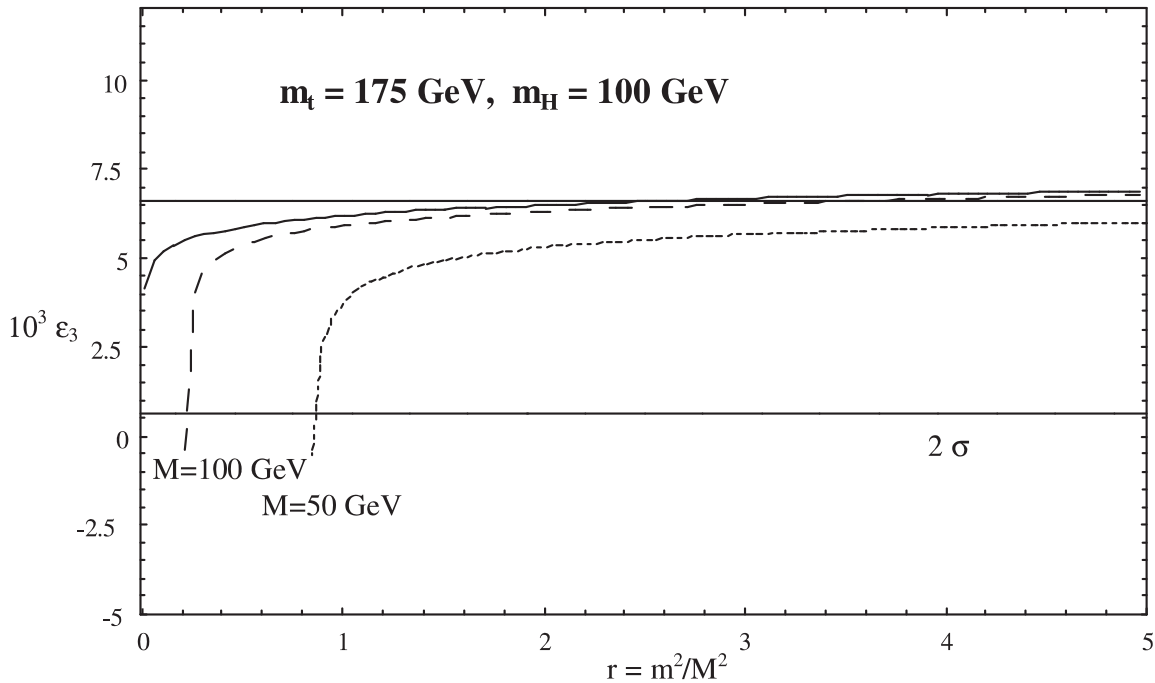,height=9cm,angle=0}}
\caption{\footnotesize\it 
Comparison between the asymptotic (solid line) and full one-loop 
(respectively with mass $M=100$ {\rm GeV} (dashed lines) and $M=50$ {\rm GeV}
(dotted lines)) computations of $\epsilon_3$ versus $r=m^2/M^2$, for an
additional quark doublet. For the SM contribution, $m_t=175$ {\rm GeV}  and
$m_H=100$ {\rm GeV} are assumed. The $2 \sigma$ allowed region is also
displayed.}
\label{pol2.eps}
\end{figure}

\subsection{Full Calculation}

We are thus lead to consider the possibility of relatively
light (but obviously under production threshold), chiral fermions, both to 
check the 
agreement with the present data, and to test the reliability of our effective 
lagrangian approach. If the additional fermions are not sufficiently heavy, 
we do not expect that their one-loop effects are accurately reproduced 
by the coefficients $a_i$ in eqs. (\ref{b1}-\ref{b2}). 
In this case we have to consider the full dependence on external momenta 
of the Green functions, not just the first two terms of the $p^2$ expansion 
given in eq. (\ref{a0}). 
We recall that in this case the $\epsilon$ parameters are given by \cite{bfc}:
\bea
\Delta \epsilon_1 &=& e_1-e_5(m_Z^2)\nn\\
\Delta \epsilon_2 &=& e_2-s^2~ e_4 -c^2 ~e_5(m_Z^2)\nn\\
\Delta \epsilon_3 &=& e_3(m_Z^2)+c^2~ e_4-c^2~ e_5(m_Z^2)
\label{a16}
\eea
where we have kept into account the fact that in our case there are no vertex 
or box corrections to four-fermion processes.
In eq. (\ref{a16})
\bea
e_1 &=& \frac{\Pi_{ZZ}(0)}{m_Z^2}-\frac{\Pi_{WW}(0)}{m_W^2}\nn\\
e_2 &=& \Pi'_{WW}(0)-c^2~\Pi'_{ZZ}(0)- 2 s c~
\frac{\Pi_{\gamma Z}(m_Z^2)}{m_Z^2}-
s^2~\Pi'_{\gamma\gamma}(m_Z^2)\nn\\
e_3(p^2) &=& \frac{c}{s}\left\{s c \left[
\Pi'_{\gamma\gamma}(m_Z^2)-\Pi'_{ZZ}(0)\right]+
(c^2-s^2)~ \frac{\Pi_{\gamma Z}(p^2)}{p^2}\right\}\nn\\
e_4 &=& \Pi'_{\gamma\gamma}(0)-\Pi'_{\gamma\gamma}(m_Z^2)\nn\\
e_5(p^2) &=& \Pi'_{ZZ}(p^2)-\Pi'_{ZZ}(0)
\label{a17}
\eea
We introduce also (for later use) $\Delta\alpha(p^2)$, 
$\Delta k(p^2)$, $\Delta\rho(p^2)$, $\Delta r_W$ and $e_6$ by the following 
relations in terms of the unrenormalized vector-boson vacuum polarizations:
\bea
\Delta\alpha(p^2) &=& \Pi'_{\gamma\gamma}(0)-\Pi'_{\gamma\gamma}(p^2)\nn\\
\Delta k(p^2) &=& -\frac{c^2}{c^2-s^2}~ (e_1-e_4)+
              \frac{1}{c^2-s^2}~ e_3(p^2)\nn\\
\Delta\rho(p^2) &=& e_1-e_5(p^2)\nn\\
\Delta r_W &=& -\frac{c^2}{s^2}~e_1+\frac{c^2-s^2}{s^2}~e_2 + 2~e_3(m^2_Z) + 
              e_4 \nn\\
e_6 &=& \Pi'_{WW}(m_W^2)-\Pi'_{WW}(0)
\label{v6}
\eea
where we have:
\be
\Pi'_{VV}(p^2) = \frac{\Pi_{VV}(p^2)-\Pi_{VV}(m_V^2)}{(p^2 -m_V^2)}~~~,
                 ~~~~~~~~~~~~~~~~~~~~~~~~~~~~~V=(\gamma, Z,W)
\ee
and finally, the effective sine is defined:
\be
s^2=\frac{1}{2}-\sqrt{\frac{1}{4}-\frac{\pi\alpha(m^2_Z)}{\sqrt{2} 
                 G_F m_Z^2}}
\ee

If $p^2=m_Z^2$ then $\Delta\alpha(p^2)$, $\Delta k(p^2)$, $\Delta\rho(p^2)$, 
$\Delta r_W$ coincides with the corrections $\Delta\alpha$, $\Delta k$, 
$\Delta\rho$ and $\Delta r_W$ which characterize the electroweak observables 
at the $Z$ resonance \footnote{
The new parameter $e_6$ is added to the five defined in \cite{bfc} for taking 
in account the presence of the 1-loop correction of the $W$'s external line 
in the $e^+ e^- \raw W^+ W^-$ process}.
The expressions for the quantities $e_i$, in the case of an ordinary quark 
or lepton doublet can be easily derived from the literature \cite{yel}. 

In fig.3 we illustrate our full one loop result in the plane
$(\epsilon_1,\epsilon_3)$, for the case of an extra quark doublet.
The upper ellipsis represents the $1 \sigma$ experimentally allowed region,
obtained by combining all LEP1 data. We plot the result for an extra quark
doublet (full line) taking $m_{t}=175$ GeV and $m_H=100$ GeV.
One of the two masses is kept fixed at $50$ GeV, and the other one runs from
$50$ GeV to $170$ GeV. One has in this way two branches, according to which
mass, $m$ or $M$, has been fixed.
If at least one of the two masses is small, this causes a substantial
deviation from the asymptotic, effective lagrangian prediction.
In particular, as it was observed in \cite{bfc}, a large negative 
contribution to both $\epsilon_1$ and $\epsilon_3$ is now possible, 
due to a formal divergence of $\Pi'_{ZZ}(m_Z^2)-\Pi'_{ZZ}(0)$ at the
threshold which produces a large and positive $e_5$. 
Clearly, this behaviour cannot be reproduced by $\LL_{eff}$, which, at the 
fourth order in derivatives, automatically sets $e_5=0$. 
The dashed line shows the predictions when all the two masses of the
additional quark doublet are heavy.
Here we fix one of the masses to $200$ GeV and let the other vary from $200$
GeV to $300$ GeV.
As before the top and Higgs masses has been fixed to $175$ GeV
and $100$ GeV. As expected, it appears that only a small amount of splitting
among the doublet components is allowed.
For the chosen value of $m_{t}$ and $m_H$, the SM prediction lies already 
outside the $1 \sigma$ allowed region and additional positive contributions to
$\epsilon_1$ tend to be disfavoured. On the contrary, the positive 
contribution to $\epsilon_3$, almost constant in the chosen range of masses,
is still tolerated.
If one also includes the SLD determination of the left-right asymmetry, 
then one gets the lower ellipsis.

\begin{figure}[t]
\vspace{0.1cm}
\centerline{
\epsfig{figure=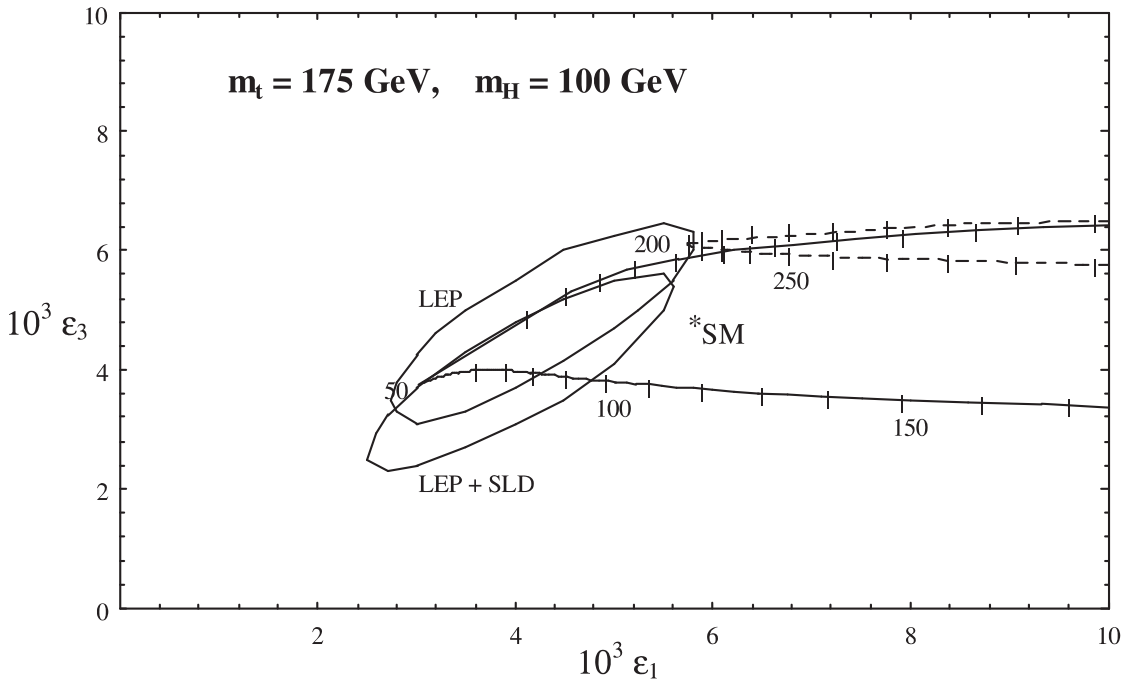,height=9cm,angle=0}}
\caption{\footnotesize\it 
Predictions for $\epsilon_1$, $\epsilon_3$ from an additional
quark doublet. The lower (upper) dashed line represents the case $m~(M)~=200$ 
GeV, $M~(m)$ varying between $200$ GeV and $300$ GeV, evaluated with 
${\LL}_{eff}$. The lower (upper) full line corresponds to $m~(M)~=50$ GeV, 
$M~(m)$ varying between $50$ GeV and $170$ GeV, evaluated with a complete 
1-loop computation. 
The SM point corresponds to $m_t=175$ GeV and $m_H=100$ GeV. The upper (lower) 
ellipses is the $1 \sigma$ allowed region, obtained by a fit of the high
energy data which excludes (includes) the SLD measurement.}
\label{pol1.eps}
\end{figure}

A relevant question is, then, when the asymptotical regime starts, i.e.
how close to $M_Z$ should be the masses of the new quarks or leptons for 
observing deviations due to the full expression of $\Pi_{ij}(p^2)$ instead 
of the truncated expression given in eq. (\ref{a0}). 
A detailed analysis shows that already for masses of the new fermions above 
$70 - 80$ GeV the difference between the values of the $\epsilon_i$ obtained 
with the truncated and full expression of $\Pi_{ij}(p^2)$ are as small as 
$10^{-4}$, i.e. below the present experimental level of accuracy. 
This is illustrated in fig.2 where the asymptotical (full line) and full one
loop (dashed and dotted) expression of $\epsilon_3$ are compared as a
function of $r$.

Beyond the indirect precision tests, the possibility of having new fermions
carrying the usual $SU(3)_C \times SU(2)_L \times U(1)_Y$ quantum numbers 
can clearly also be bounded by the direct searches.
Concerning the present searches, from LEP1 we have the lower bound of $M_Z/2$
\footnote{Recent analysis \cite{lep15} of data taken during the upgrade from 
LEP1 to LEP2 increase the lower bound for fermion masses from $M_Z/2$ 
to $\simeq 60$ GeV.
} 
which applies independently from any assumption on the decay modes of the new 
fermions which couple to the $Z$ boson. Much stronger limits on the new quark 
masses can be inferred from the Tevatron results. However, as we know from the
search for the top quark, these latter bounds rely on assumptions concerning
the decay modes of the heavy quark. For instance, in the case of the top
search it was stressed that if a new decay channel into the $b$ quark and a
charged Higgs were avaible to the top, then one could not use the CDF
bounds on $m_t$ \cite{cdf} which came along these last years, before the
final discovery of the top quark.

Now, it may be conceivable that the new physics related to the presence of
extra-fermions can also affect their possible decay channels making the
lightest of the new fermions unstable.
Indeed, we stated in our assumption that the new fermions do not
essentially mix with the ordinary ones, hence one has to invoke new physics
if one wants to avoid the formation of stable heavy mesons made out of the
lightest stable new fermion and of the ordinary fermions of the Standard
Model. If the new fermions can decay within the detector, then the bounds on 
their masses, coming from Tevatron data, must be discussed in a model-dependent
way and even the case of new quarks with masses lighter than $m_t$ are not 
fully ruled out.

If on the contrary the lightest new quark is stable, then searches for exotic
heavy meson at CDF already ruled out the possibility of being near the
threshold $M_Z/2$. The existence of coloured particle with charge $\pm 1$
is strictly bounded over $130 \ GeV$ from CDF experiment \cite{sta}.
Finally, note that for charged leptons the bound coming from CDF are much
less stringent. A new stable charged lepton of mass of $60 - 70 \ GeV$ cannot
be ruled out.


\section{Three Points Functions}

If new physics beyond the SM were modeled by additional heavy chiral fermions,
of the kind we have considered, then we could draw informations on the 
searches, at future colliders, of anomalous trilinear gauge boson couplings 
of the $WWV$ vertex (where $V$ stands for neutral vector boson). 
We define the kinematics of the $WWV$ vertex as
\be
V(p,v_1) \raw W^-(q,v_2) + W^+(\bar q,v_3)~~~~~,
\ee
where $p, q, \bar q$ are the momenta of $V, W^-, W^+$ respectively, and 
$v_1, v_2, v_3$ are their polarization vectors. For simplicity we can impose 
the produced $W^\pm$ to be on shell, so:
\be
q^2=\bar{q}^2=m_W^2~~~~,~~~~q \cdot v_2 = \bar q \cdot v_3 = 0~~~~~.
\ee
Following the definitions of \cite{hpz}, the general CP-conserving coupling 
of two on-shell charged vector bosons with a neutral vector boson 
($V= \gamma, Z$) can be derived from the following effective lagrangians:
\bea
\frac{{\cal L}_{WWV}}{g_{WWV}} &=& i g_1^V (W_{\mu \nu}^{\dagger} W^\mu V^\nu -
 W_{\mu}^{\dagger} V ^\nu W_{\mu \nu}) + 
i \kappa_V W_{\mu}^{\dagger} W^\nu V^{\nu \mu}\nn \\
 & + & i \frac{\lambda_V}{\Lambda^2} W_{\lambda \mu}^{\dagger} W^{\mu}_{\nu} 
          V^{\nu \lambda} + g_5^V\epsilon^{\mu\nu\rho\sigma}
(W_{\mu}^{\dagger}\stackrel{\lraw}{\partial}_{\rho} W_\nu) V_\sigma~~~~,
\label{leff}
\eea
where $V_\mu$ is the neutral vector boson field, $W_\mu$ is the field 
associated 
with the $W^-$, $W_{\mu \nu}=\partial_\mu W_\nu -\partial_\nu W_\mu$,  
$V_{\mu \nu}=\partial_\mu V_\nu - \partial_\nu V_\mu$ and $\Lambda$ is a mass 
scale parameter opportunely chosen.
By convention $g_{WW\gamma}=-e$ and $g_{WWZ}=-e~c/s$. 

In momentum space the $WWV$ vertex can be decomposed as
\bea
\Gamma_V^{\alpha\beta\mu}(q,\bar q,p) &=& f_1^V (q - \bar q)^\mu 
g^{\alpha\beta}-\frac{f_2^V}{m_w^2} (q-\bar q)^\mu p^\alpha p^\beta + 
f_3^V (p^\alpha g^{\mu \nu} - p^\beta g^{\alpha\mu})\nn \\
         &+& i f_5^V \epsilon^{\alpha \beta \mu \sigma}(q-\bar q)_\sigma.
\label{veff}
\eea  
Here all the forms factors $f_i^V$ are dimensionless functions of $p^2$.
From eqs. (\ref{leff}-\ref{veff}) it is easy to recover the following 
relations:
\bea
f_1^V &=& g_1^V + \frac{p^2}{2 \Lambda^2} \lambda_V\nn\\
f_2^V &=& \lambda_V\nn\\
f_3^V &=& g_1^V + \kappa_V + \lambda_V\nn\\
f_5^V &=& g_5^V
\label{feff}
\eea
Obviously we can add to these effective lagrangians ${\cal L}_{WWV}$ higher 
dimension operators, by replacing $V_\mu$ by $\partial^{2n} V_\mu$ 
(with n arbitrary integer).
Higher order operators in eq. (\ref{leff}) will contribute with $p^{2n}$ terms 
to the right side of eq. (\ref{feff}).
We need only the 4 form factors of eq. (\ref{veff}) for 
parametrizing all the new physics effects in the trilinear gauge effects.

If we are working a tree level all is univocally defined.
But if we are adding also one loop contributions, one has to declare exactly 
which contributions wants to include in the "form factors" definitions. 
Defining a renormalization depending quantity ${\cal K}$ we can write the 
forms factors as:
\be
f^V_i = f^{SM}_i + \Delta f^V_i~~~~~~,~~~~~~\Delta f^V_i = 
{\cal K}\, f^{SM}_i + \delta f^V_i~~~~~~(i=1,2,3,5).
\label{frin}
\ee
where $\Delta f^V_i$ denote the full one loop contributions due to new physics 
virtual effects, and $\delta f^V_i$ is the pure trilinear contributions.
The term ${\cal K}\, f^{SM}_i$ depends on the choice of the overall normalization 
of the trilinear vertex $WWV~~(V=\gamma,Z)$, usually denoted by $g_{WWV}$, and 
on the renormalization scheme adopted.
To avoid this indetermination we prefer to define these couplings by observing 
the physical process of $W^\pm$ pair production in $e^+ e^-$ collision,
\be
e^-(k,\sigma)~~+~~e^+(\bar{k},\bar{\sigma})~~\raw~~W^-(q,\lambda)~~+~~
W^+(\bar{q},\bar{\lambda}) 
\ee
Following \cite{hpz} the helicity amplitude for this process can be written
\be
{\cal M} = \sqrt{2}~e^2~\M2 (\Theta)~\epsilon~ 
            d^{J_0}_{\Delta\sigma,\Delta\lambda}(\Theta)
\ee
where $\epsilon = (-1)^\lambda \Delta\sigma$ is a sign factor, 
$\Delta\sigma=\sigma-{\bar\sigma}$, $\Delta\lambda=\lambda-{\bar\lambda}$, 
$J_0=max(|\Delta\sigma|,|\Delta\lambda|)$, $\Theta$ is the scattering angle of 
$W^-$ with respect to $e^-$ in the $e^+e^-$ c.m. frame and 
$d^{J_0}_{\Delta\sigma,\Delta\lambda}$ are angular functions depending from 
the helicity of the initial and final states.

After the inclusion of the 1-loop corrections due to the heavy fermions
and of the appropriate counterterms, the reduced amplitude for the 
process at hand reads
\noindent
\bea
\bullet \Delta\lambda=\pm2 & &\nn\\
\M2 & = & -\frac{\sqrt{2}}{s^2}\frac{\delta_{\Delta\sigma,-1}}
{1+\beta^2-2 \beta \cos\Theta}\left[1-\frac{s^2}{c^2-s^2}
\Delta r_W-e_6\right]
\label{v2}
\eea
\bea
\bullet |\Delta\lambda| \le 1 & & \nn\\
\Mg &=& -\beta \delta_{|\Delta\sigma|,1} \left[1+\Delta\alpha(p^2)\right]~
                            \left[\Asm  + \DAg \right]\nn\\
\Mz &=&\beta~\frac{p^2}{p^2-m_Z^2}~\left[\delta_{|\Delta\sigma|,1}-
\frac{\delta_{\Delta\sigma,-1}}{2s^2 (1+\Delta k(p^2))} \right]\nn\\
 && ~~~~~~~~~~~~~~~~~~~~~~~\left[1+\Delta\rho(p^2)+\frac{c^2-s^2}{c^2}
\Delta k(p^2)\right]~\left[\Asm  + \DAz \right]\nn\\
\Mnu&=& \frac{\delta_{\Delta\sigma,-1}}{2s^2~\beta}  
\left[ 1 - \frac{s^2}{c^2-s^2} \Delta r_W-e_6 \right]
\left[ B^{SM}_{\lambda{\bar\lambda}} - \frac{1}{1+\beta^2-2 \beta \cos\Theta} 
       C^{SM}_{\lambda{\bar \lambda}}\right]
\label{v3}
\eea
with $\beta=(1-4 m_W^2/p^2)^{1/2}$.
$\Asm$, $B^{SM}_{\lambda{\bar\lambda}}$ and $C^{SM}_{\lambda{\bar\lambda}}$ 
are the tree-level SM coefficients listed in Tab.1. 
\begin{table}[t]
\centering
\begin{tabular}{||c|c|c|c||} \hline
& & & \\
$\lambda\bar\lambda$ & $\Asm$ & $B^{SM}_{\lambda \bar\lambda}$ & 
                               $C^{SM}_{\lambda \bar\lambda}$ \\
& & & \\ \hline
$++ \, ,\,-- $ & $ 1 $ & 1 & $1/ \gamma^2$ \\
$+0 \, ,\,0-$ & $ 2 \gamma $ & $2\gamma$ & $2(1+\beta)/\gamma$ \\
$0+ \, ,\,-0$ & $2 \gamma $ & $2\gamma$ & $2(1-\beta)/\gamma$ \\
$00$ & $2 \gamma^2 + 1$ & $2\gamma^2$ & ${2/ \gamma^2}$ \\ \hline
\end{tabular}
\caption{\footnotesize\it Standard Model coefficients expressed in terms of 
$\gamma^2=p^2/4 m_W^2$.}
\end{table}
The coefficients $\DAg$ and $\DAz$  can be expressed in terms of the 
$CP$-invariant form factors according to the relations:
\bea
\Delta A_{++}^V & = &\Delta A_{--}^V = \Delta f_1^V\nn\\
\Delta A_{+0}^V & = &\Delta A_{0-}^V = \gamma(\Delta f_3^V-i\Delta f_5^V)\nn\\
\Delta A_{-0}^V & = &\Delta A_{0+}^V = \gamma(\Delta f_3^V+i\Delta f_5^V)\nn\\
\Delta A_{00}^V & = & \gamma^2\left[-(1+\beta^2)\Delta f_1^V + 4 \gamma^2 
                       \beta^2 \Delta f_2^V + 2\Delta f_3^V\right]
\label{v4}
\eea
Hence our convention for the form factors consist in taking 
${\cal K}= \Pi'_{WW}(m^2_W)$ in eq. (\ref{frin}), and consequentially 
$\Delta f_i^V \equiv \Pi'_{WW}(m^2_W) f_i^{SM} + \delta f_i^V$.
So $\Delta f_i^V~~(i=1,2,3,5)~~(V=\gamma,Z)$ includes both
the contribution coming from the 1-loop correction to the vertex $WWV$
and the wave-function renormalization of the external $W$ legs, taken on the 
mass-shell. This makes the terms $\Delta f_i^V$ finite.

Finally, it's worth make few comments about the unitarity constrains. 
In the high-energy limit, the individual SM amplitudes from photon, 
$Z$ and $\nu$ exchange are proportional to $\gamma^2$ when both the 
$W$ are longitudinally polarized ($LL$) and proportional to $\gamma$ when 
one $W$ is longitudinal and the other is transverse ($TL$). 
The cancellation of the $\gamma^2$ and $\gamma$ terms in the overall 
amplitude is guaranteed by the tree-level, asymptotic relation 
$\Asm~=~B^{SM}_{\lambda{\bar\lambda}}$.
When one loop contributions are included, one has new terms proportional 
to $\gamma^2$ and $\gamma$ (see $\DAg$ and $\DAz$ in eq.(\ref{v4})) and the 
cancellation of those terms in the high-energy limit entails relations among 
oblique and vertex corrections.
Omitting, for instance, the gauge boson self-energies such cancellation does 
not occur any longer and the resulting amplitudes violate the requirement of 
perturbative unitarity. 
So the way it happens shows us the relevance of considering both the 
bilinear and trilinear contributions in the results of eq. (\ref{v3}).

On the other hand, one of the possibilities to have appreciable deviations
in the cross-section is to delay the behaviour required by unitarity.
This may happen if in the energy window $m_W << \sqrt{p^2}\le 2 M$
($M$ denoting the mass of the new particles)
the above cancellation is less efficient and terms proportional
to positive powers of $\gamma$ survive in the total amplitude.
If $\gamma$ is sufficiently large, which is not the case for LEP2,
then a sizeable deviation from the SM prediction is not unconceivable.
It's useful to introduce the following quantity
\be
\Delta R_{AB}=\frac{\dd\left(\frac{d\sigma}{d\cos\Theta}\right)_{AB} -
\dd\left(\frac{d\sigma}{d\cos\Theta}\right)^{SM}_{AB}}
{\dd\left(\frac{d\sigma}{d\cos\Theta}\right)^{SM}_{AB}} ~~~~~~ {\rm with}
~~~~~~\frac{d\sigma}{d\cos\Theta} = \frac{\beta}{32~\pi~p^2} |{\cal M}|^2
\label{v9}
\ee
representing the relative deviation from $SM$ results due to New Physics
effects in the different helicity channels $AB = LL,TL,TT,tot$.

\subsection{Static Approximation}

As before, in this section, we are interested only in the low-energy 
($p^2 \ll M^2$) process.
In this limit we can chose $\Lambda^2=M^2$ and neglect the term
$\lambda_V/\Lambda^2$ in eq.(\ref{leff}).

From the effective lagrangian of eqs. (\ref{a9}-\ref{b2}) the anomalous 
trilinear couplings can be expressed as combinations of the coefficients
$a_i$ and the renormalization depending quantity ${\cal K}^V$:
\bea
\Delta f_1^\gamma &=& \Delta f_5^\gamma = {\cal K}^\gamma~~~~,\nn \\
\Delta f_3^\gamma &=& - g^2(a_1 - a_2 + a_3 - a_8 + a_9)+2\,{\cal K}^\gamma
                        ~~~~,\nn \\
\Delta f_1^Z &=& -\frac{g^2}{c^2} a_3 + {\cal K}^Z~~~~,\nn \\
\Delta f_3^Z &=& g^2\left[\frac{s^2}{c^2}(a_1 + a_{13} - a_2) - a_3 + a_8 -
                           a_9 + a_{13}\right] + 2\,{\cal K}^Z~~~~,\nn \\
\Delta f_5^Z &=& \frac{g^2}{c^2} a_{14}~~~~~,
\label{a15bis}
\eea
In in the static approximation limit ${\cal K}^V = \Pi'(m^2_W) = 0$, so 
$\Delta f_i^V = \delta f_i^V$.
These formulas can be readily evaluated by substituting in eq.(\ref{a15bis})
the explicit expressions of the coefficients $a_i$ given in eqs. 
(\ref{a9}-\ref{b2}).

As suggested by the small allowed value of the $\epsilon_1$ parameter, if we
restrict our analysis in particular to the case of degenerate quark doublet
\footnote{For a leptonic doublet the contributions in the $r=1$ limit are exactly 
$1/3$ of the hadronic one.} we find very small values for the form factors:
\bea
\delta f_1^\gamma &=& \delta f_5^\gamma = 0~~~~~,\nn \\
\delta f_3^\gamma &=& -\frac{G m_W^2}{4\pi^2\sqrt{2}}\sim - 1.3\cdot 10^{-3}
                        ~~~~~,\nn \\
\delta f_1^Z &=& -\frac{G m_W^2}{4\pi^2\sqrt{2}}~\frac{1}{c^2} 
                  \sim - 1.7 \cdot 10^{-3}~~~~~,\nn \\
\delta f_3^Z &=& -\frac{G m_W^2}{4\pi^2\sqrt{2}}~\frac{1+c^2}{c^2}
                  \sim - 3.1\cdot 10^{-3}~~~~~,\nn \\
\delta f_5^Z&=& 0~~~~~.
\eea
Obviously other conventions are possible, which give different expressions for 
the form factors. For example putting  
\be
{\cal K}^Z =  
\frac{1}{c^2-s^2}~\left[a_0+\frac{e^2}{c^2}(a_1+a_{13})\right]~~~~,~~~~
{\cal K}^\gamma = 0~~~~~, 
\ee
we obtain the relations found by \cite{par,app,fmrs}.

In the $m^2_W \ll p^2 \ll M^2$ limit, and for $\cos\Theta \ll 1$\footnote{ 
Only in this region we can safely neglect the contributions from the 
$C^{SM}_{\lambda \bar\lambda}$ terms, that Tab.1 shows are of the order 
$1/\gamma$ or $1/\gamma^2$.}, 
we derived the analytical expression for $\Delta R_{LL}$
\be
\Delta R_{LL} = \frac{g^2}{16 \pi^2}~4~\gamma^2 = 
                 \frac{4 \sqrt{2} G}{16 \pi^2}~p^2~~~~~. 
\label{drstatico}
\ee
Hence the deviation from the SM grows like $p^2$. But this not represents an 
unitarity violation, because must be always satisfies the requirement 
$p^2 \ll M^2$. 
Even if this is an extremely simplified formula we obtain a realistic 
indication on the dimension of the effects we are playing with. Putting 
$\sqrt{p^2}=500$ GeV in eq. (\ref{drstatico}) we have $\Delta R_{LL} \sim 
0.10$, very close to the exact value of fig. \ref{pol3}, calculated for the 
same energy and with $M=1000$ GeV, in the region $\cos\Theta \ll 1$.   
Similar $\gamma^2$ dependence can also be derived for $\Delta R_{LT}$, while 
the deviations from the SM values for $\Delta R_{TT}$ are of the order
$\gamma^0$, because either the SM  and New Physics contributions have no
dependence from positive powers of $\gamma$ (see Tab.1 and eq.(\ref{v4})).

\subsection{Full Calculation}

The results obtained in the previous sub-section, whenever suggestive, are
not completely satisfactory, essentially for two reasons:
\begin{itemize}
\item
We learned from the bilinear analysis that in some regions of the phase space,
i.e. near the production threshold, the effective lagrangian becomes
unreliable and effects of the order of $p^2/M^2$ can not be neglected.
\item
The approximations that are behind the derivation of eq.(\ref{drstatico}) loose 
sense in two, in principle very important, regions:
\begin{itemize}
\item for $\sqrt{p^2} = 200$ GeV (LEP2), where $m_W$ is no more negligible 
respect to the c.m. energy
\item for $\sqrt{p^2} = 1000$ GeV (NLC), where the mass of the extra virtual 
fermion is required to be of the order of the energy by the stability of the 
Higgs potential (for example \cite{fmm} put as upper limit $M \sim 1.5$ TeV).
\end{itemize} 
\end{itemize} 
 
So we are lead to analyse the full one loop calculation, having in mind two
complementaries aspects:
\begin{itemize}
\item we want to inspect how far from the threshold we can push the exclusion 
of extra fermion doublets,
\item we want to know how grow these effects with growing energies.
\end{itemize}

The general expression for the form factors are rather complex. A simplified 
expression for $\Delta R_{LL}$ can be derived again for $\cos \Theta \ll 1$, 
and in the limit $m_W \ll M~,~\sqrt{p^2}$:
\be
\Delta R_{LL}= \frac{g^2}{16 \pi^2}~\frac{4 M^2}{p^2}~4 \gamma^2
 \cdot N_c F\left(\frac{4 M^2}{p^2}\right) 
\label{v10}
\ee
with the function $F$ defined:
\be
F\left(x \right) = \left[1-\sqrt{x-1}~\arctan\frac{1}{\sqrt{x-1}}\right]
\ee
For $p^2>>M^2$, $F$ grows only logarithmically and unitarity is 
respected.
When $M^2>>p^2$, $F\simeq p^2/12 M^2$ and the decoupling property is
violated, as one expects in the case of heavy chiral fermions.
In the range of energies we are interested in, $m_W<<\sqrt{p^2}\le 2M$,
$\Delta R_{LL}$ is of order $G_F M^2$, which explains the magnitude of the 
effect exhibited in fig.\ref{pol3}. 
A similar behaviour is also exhibited by the $TL$ channel.

The full calculation results are showed in fig. \ref{pol3} for the case 
of a degenerate quark doublet. 
We plot the relative deviation from SM results \cite{arg} relative to the 
$LL$ channel, as function of $\cos \Theta$ at $\sqrt{p^2} = 200~,~500~,~1000$ 
GeV for several values of the fermions mass $M = 150~,~300~,~600~,~1000$ Gev. 
At LEP2 energy the deviations in the LL channel are of the order of 
$1-2$ per cent either for relatively light and heavy fermions.
Only when one consider particles very closed to the production threshold, 
deviations of some per-cent are achieved. In any case well below the 
observability level, because the number of events in the LL (or even the TL ) 
channel will be very small at the foreseen luminosity. 
On the other side in the $TT$ channel, where instead the number of events 
will be "adequate", the deviations expected are not enhanced by the $\gamma$ 
factor and stay at the order of $0.1$ percent.  

More interesting will be the situation at the higher energies of NLC (
$\sqrt{p^2}=500~,~1000$ GeV), in which, as one can easily see in eq.(\ref{v10}) 
a delay of unitarity in the LL (LT) channels, due to $\gamma$ enhancement 
factor, gives deviations from SM of the order 10-50 per cent, for a wide 
range of new particles masses. The effects at $\sqrt{p^2}=1000$ GeV become soon
very large reaching the limit of the validity of the perturbation expansion.
In the right-bottom graphics of fig.\ref{pol3} we summarize the energy 
behaviour for a fixed mass $M=1000$ GeV doublet.
          
\begin{figure}[t]
\vspace{0.1cm}
\centerline{
\epsfig{figure=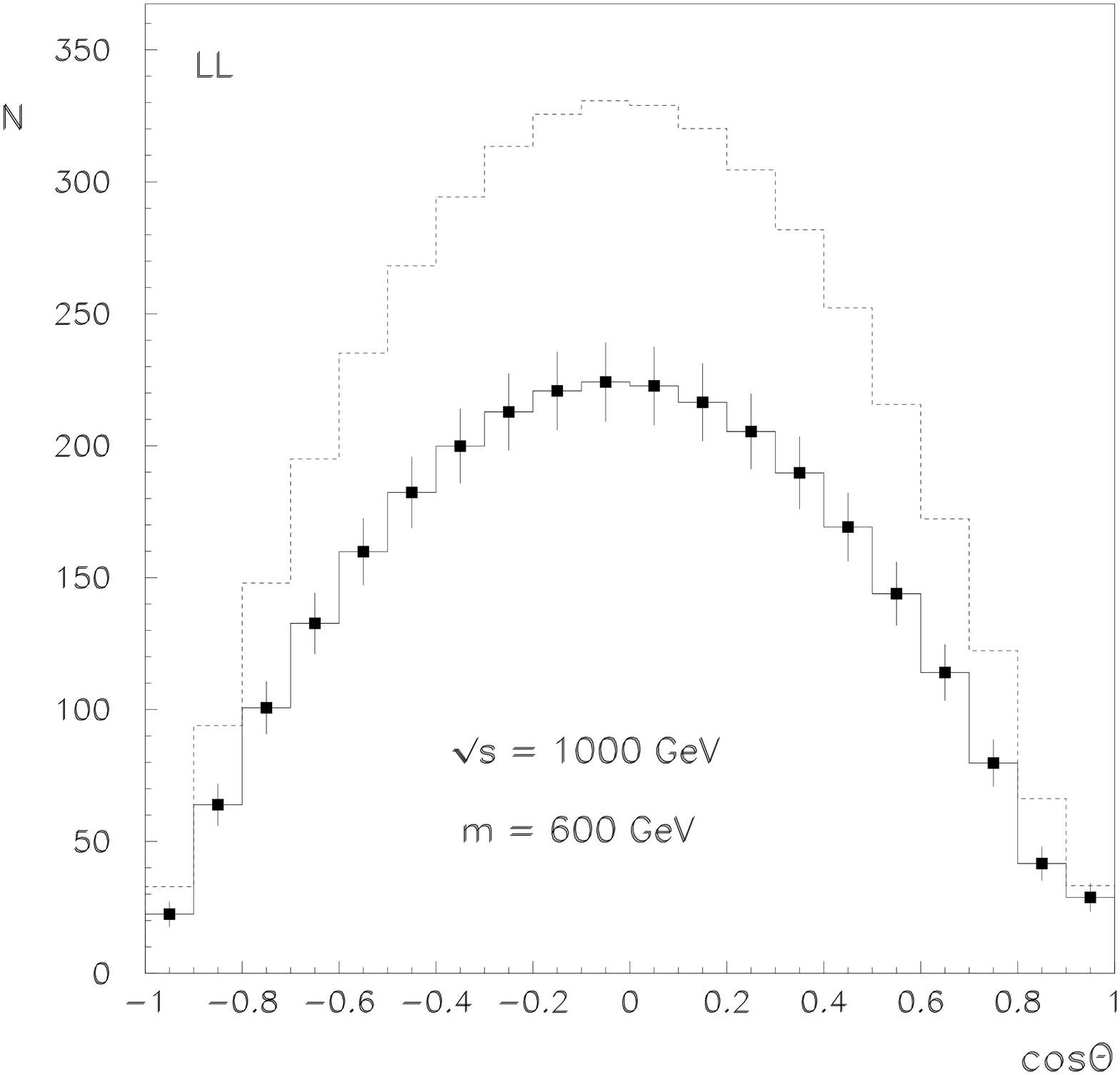,height=9cm,angle=0}}
\caption{\footnotesize\it 
Number of events predicted from the SM (full line) at energy $p^2=1000^2$ Gev
Gev and luminosity $L=100\,fb^{-1}$ the number expected for one extra doublet 
of heavy chiral fermion (dotted line) of mass $M=600$ GeV. 
The expected statistical errors are also shown.}
\label{pol4.eps}
\end{figure}

Also if the differential cross section in the LL channel is two order below 
the TT one, with the energies and the luminosities promised by NLC this effects 
will be easily seen. 
In fig.\ref{pol4.eps} we plot the number of events per bin 
at $\sqrt{p^2}=1000~GeV$ for channel LL, taking $M=600~GeV$
and assuming a luminosity of $100~fb^{-1}$. 
The error bars refer to the statistical error, dots denote the SM
expectations and the full line is the prediction for one extra heavy chiral 
fermion doublet.
We notice a clear indication of a significant signal, fully
consistent with present experimental bounds.

\begin{figure}[t]
\begin{tabular}{lr}
\epsfig{figure=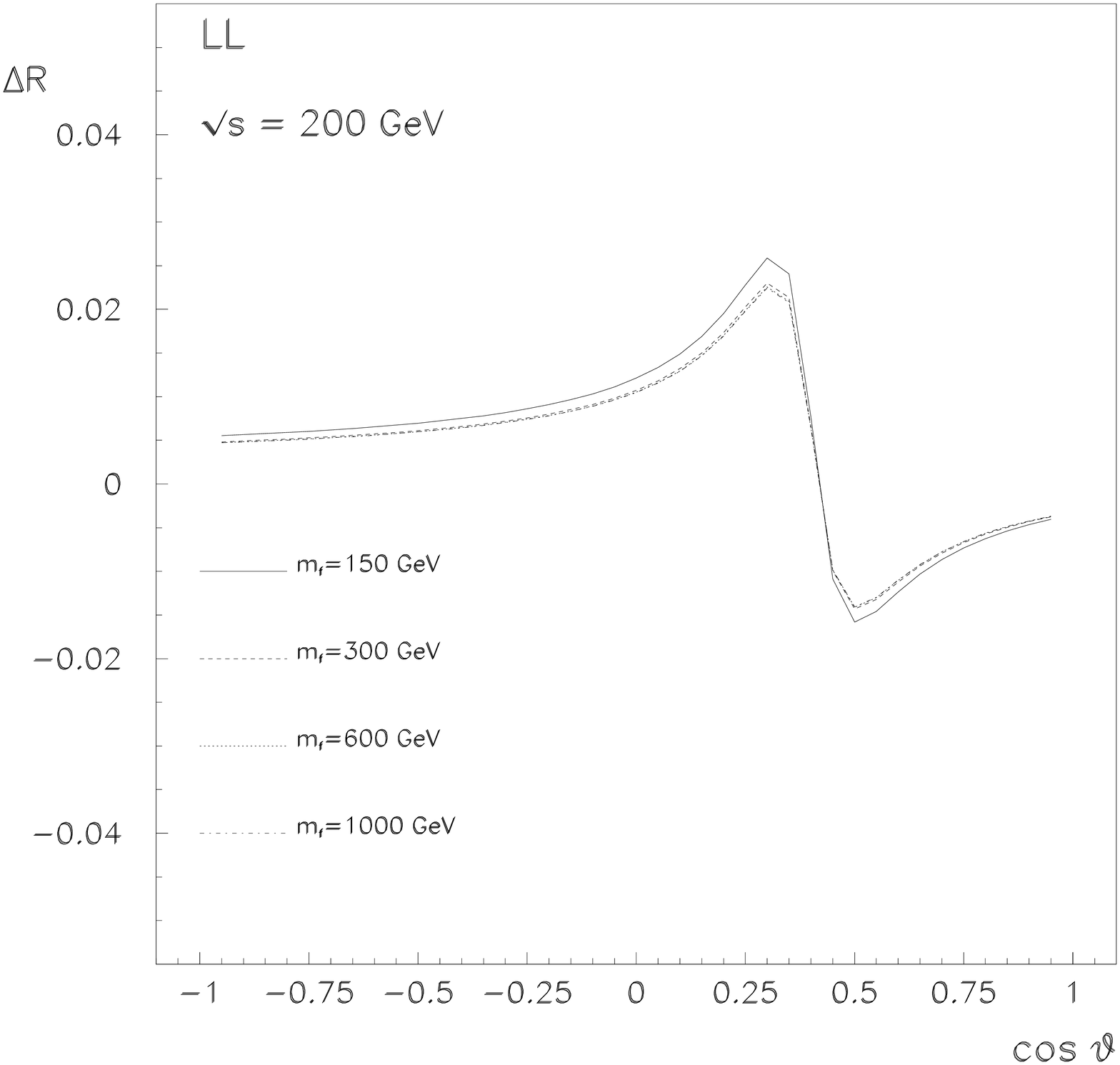,height=8.5cm,angle=0} & 
\epsfig{figure=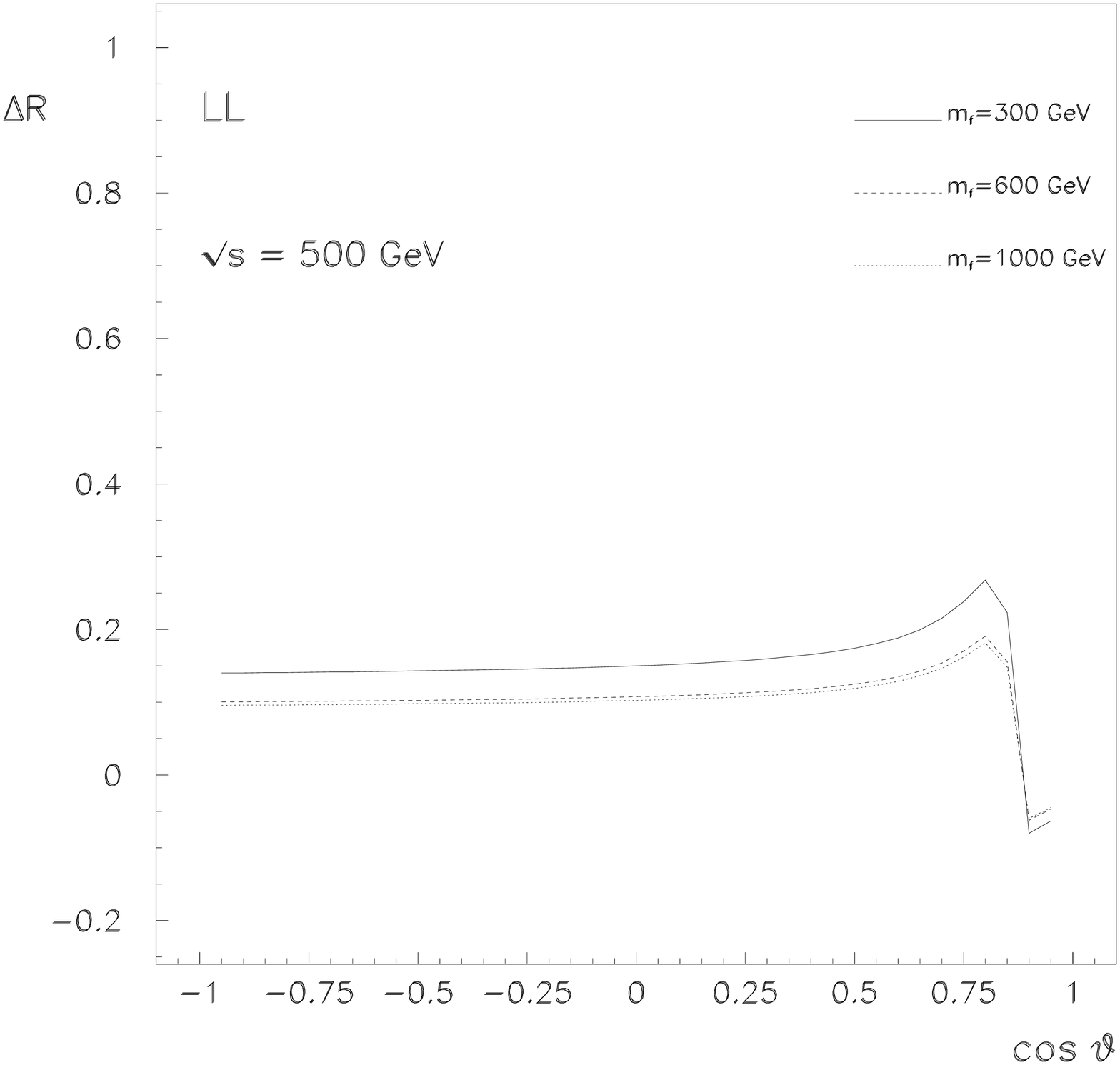,height=8.5cm,angle=0} \\
\epsfig{figure=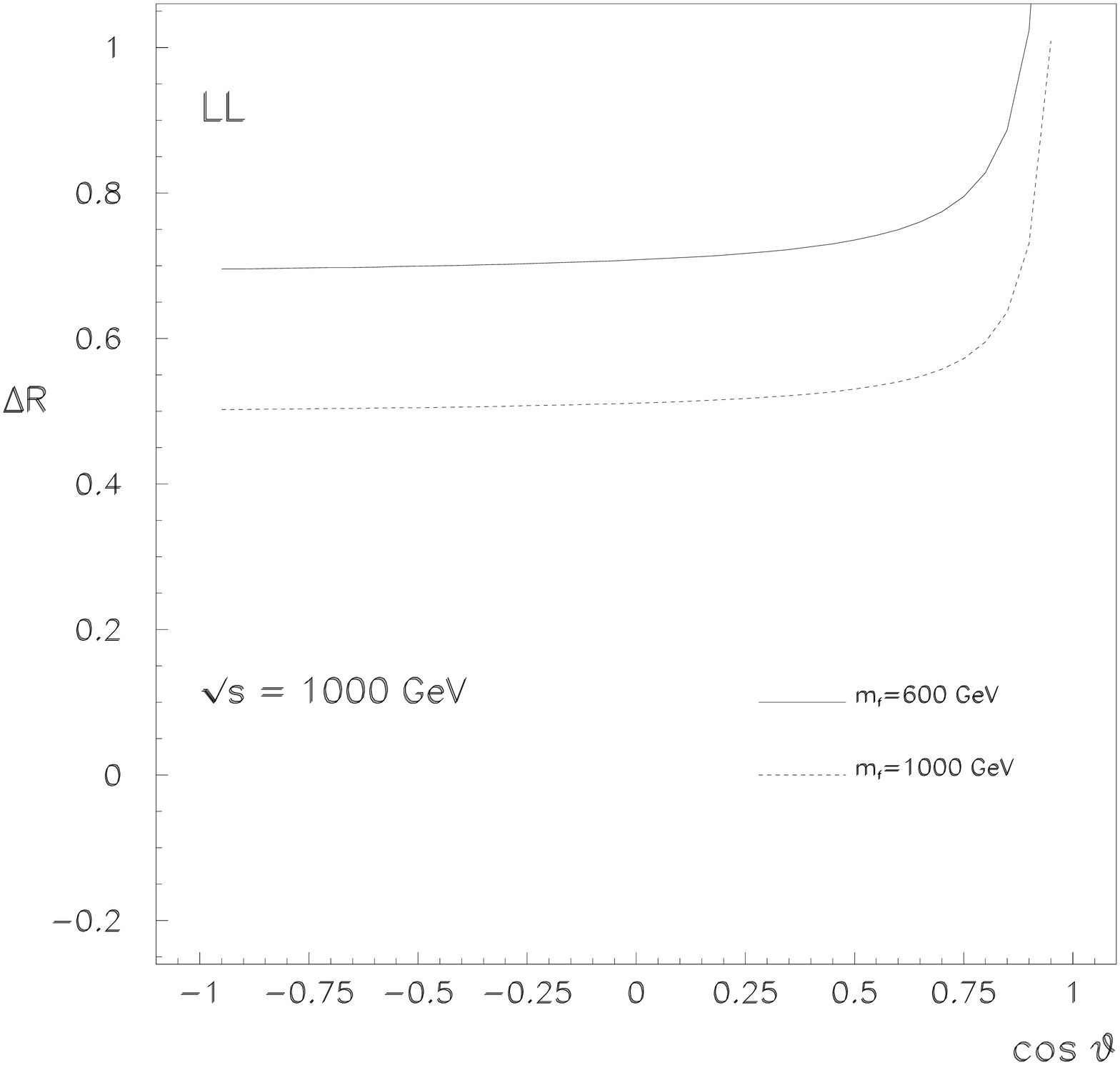,height=8.5cm,angle=0} & 
\epsfig{figure=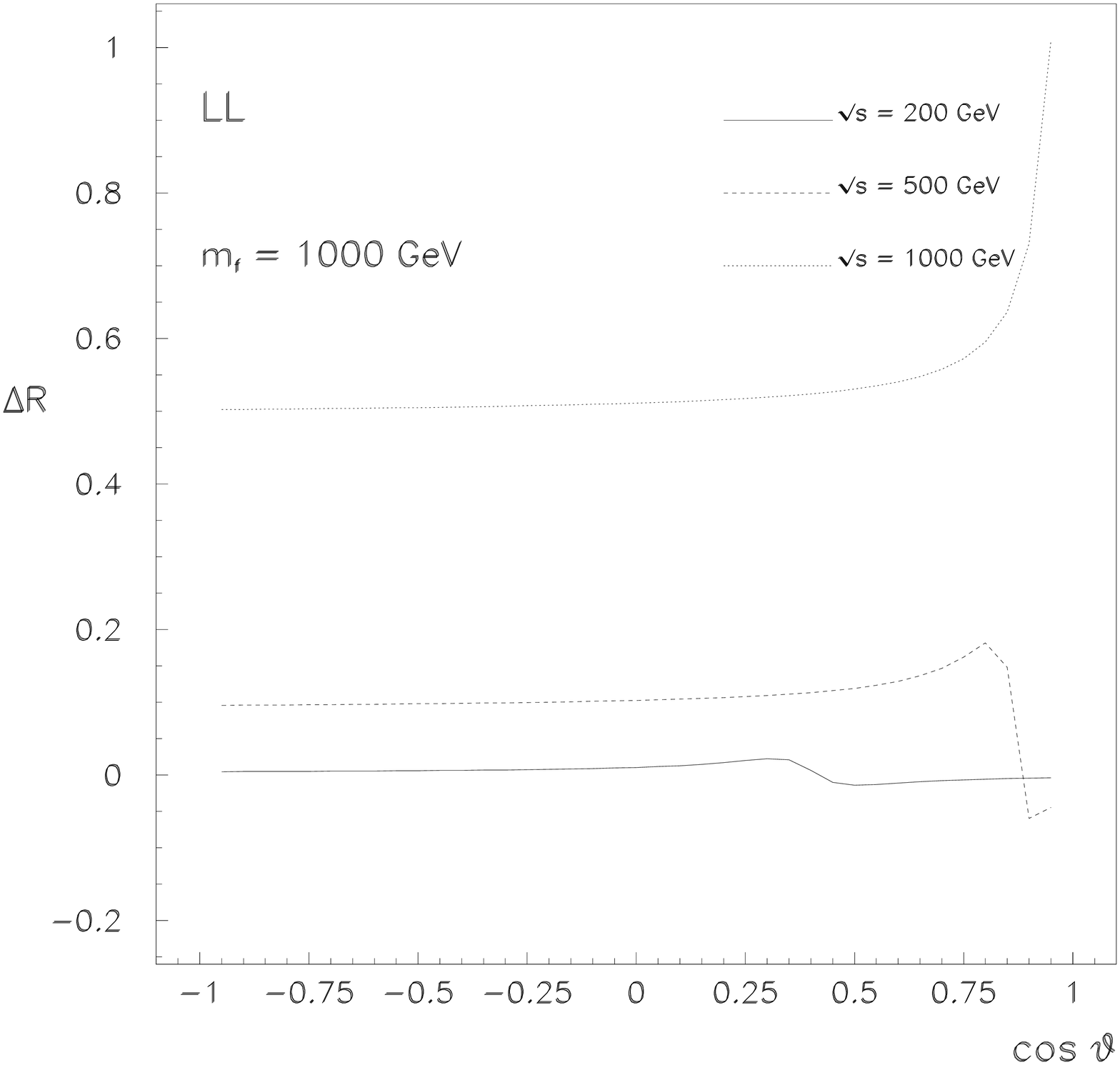,height=8.5cm,angle=0} \\
\end{tabular}
\caption{\footnotesize\it 
Relative deviations in the differential cross section due to heavy fermion 
contribution ($LL$ channel) for different c.m. energies and different masses.}
\label{pol3}
\end{figure}

Finally we would like to mention that this behaviour is typical only for 
heavy chiral fermions. We checked explicitly that, in the case of vector-like 
fermions (like for example the MSSM ones in some simplified case), no 
unitarity delay takes place also at high energies ( $\sqrt{p^2}=500$ GeV or 
even more). The deviations $\Delta R$ remain under the percent level, making 
questionable the possibility of observing such effect also in next generation 
$e^+ e^-$ colliders \cite{ls}.

\section{Conclusion}

In conclusion, we discussed the impact of the presence of new sequential
fermions on the electroweak precision tests. We showed that the present data
still allow the presence of a new quark and/or lepton doublet with masses
greater than $M_Z/2$.
Only for light new fermions which are close to the threshold $M_Z/2$ one finds
drastic departures of the effective lagrangian result from the full one-loop 
radiative corrections obtained in SM. 
The presence of new fermions carrying usual $SU(3)_C \times SU(2)_L \times 
U(1)_Y$ quantum numbers with mass as low as $60-80 \ GeV$ is severely 
limitated both by accelerator results and cosmological constraints. 

For heavier chiral fermions, at the energies provided by the NLC, a huge 
effect, due to the $\gamma$ enhancement factor, connected with a delay of 
unitarity shows in the LL, LT channels  deviations from SM of the order 10-50 
per cent, for a wide range of new particles masses. These effects seem to be 
easily measured at the hoped luminosities of NLC.  


\section*{Acknowledgements}
I'm indebted to A. Masiero, F. Feruglio and R. Strocchi, for the very pleasent 
collabaration on which this lectur is based.
I would like also to thank G. Degrassi, A. Culatti  and A. Vicini 
for the helpful discussions and suggestions.
A special thank to the organizers of the School J. Polak, J. Sladkowski and 
M. Zralek, as well to all the partecipants for the very nice hospitality 
enjoyed in Bystra.

\end{document}